\documentclass[twocolumn,epjc3]{svjour3}          

\RequirePackage[T1]{fontenc}
\smartqed  

\RequirePackage{graphicx}
\RequirePackage{epstopdf}
\RequirePackage[colorlinks,citecolor=blue,urlcolor=blue,linkcolor=blue]{hyperref}
\RequirePackage{amssymb}
\RequirePackage{slashed}
\RequirePackage{amsmath}
\hypersetup{pdfstartview=}
\journalname{Eur. Phys. J. C}

\makeatother

\begin{document}

\title{New estimate of the chromomagnetic dipole moment of quarks in the standard model}

\author{A. I. Hern\'andez-Ju\'arez    \thanksref{e1,addr1}\and
        A. Moyotl  \thanksref{e2,addr2}\and
        G. Tavares-Velasco  \thanksref{e3,addr1}
}

\thankstext{e1}{alaban7\_3@hotmail.com}
\thankstext{e2}{agustin.moyotl@uppuebla.edu.mx}
\thankstext{e3}{gtv@fcfm.buap.mx}

\institute{Facultad de Ciencias F\'isico-Matem\'aticas,\\
  Benem\'erita Universidad Aut\'onoma de Puebla,\\
 C.P. 72570, Puebla, Pue., Mexico \label{addr1} \and Ingenier\'ia en Mecatr\'onica,\\ Universidad Polit\'ecnica de Puebla,\\ Tercer Carril del Ejido Serrano s/n, San Mateo Cuanal\'a, Juan C. Bonilla,\\ Puebla, Puebla, M\'exico \label{addr2}}

\date{Received: date / Accepted: date}

\maketitle
\begin{abstract}
A new estimate of the one loop contributions of the standard model to the  chromomagnetic dipole moment (CMDM) $\hat \mu_q(q^2)$ of quarks is presented with the aim to  address a few disagreements arising in previous calculations. The most general case with arbitrary $q^2$ is considered and analytical results are obtained in terms of Feynman parameter integrals and Passarino-Veltman scalar functions, which are then expressed in terms of closed form functions when possible.
It is found that while  the QCD contribution to the static CMDM ($q^2=0$) is infrared divergent, which agrees with previous evaluations and stems from the fact that this quantity has no sense in perturbative QCD, the off-shell  CMDM ($q^2\neq0$) is  finite and gauge independent, which is verified  by performing the calculation for arbitrary gauge parameter $\xi$ via both a renormalizable linear $R_\xi$ gauge and the background field method. It is thus argued that the off-shell $\hat \mu_q(q^2)$  can  represent a valid observable quantity. For the numerical analysis we consider the region 30 GeV$<\|q\|<$ 1000 GeV and analyze the behavior of $\hat \mu_q(q^2)$ for all the standard model quarks. It is found that the CMDM of light quarks is considerably smaller than that of the  top quark as it is directly proportional to the quark mass. In the considered energy interval, both the real and imaginary parts of $\hat\mu_t(q^2)$ are of the order of $10^{-2}-10^{-3}$, with the largest contribution arising from the QCD induced diagrams, though around the threshold $q^2=4m_t^2$ there are also important contributions from diagrams with $Z$ gauge boson and Higgs boson exchange.

\end{abstract}


\section{Introduction}
The anomalous magnetic dipole moment (MDM) of fer\-mions has been a fertile field of study, giving rise to a plethora of theoretical work   within the context of the standard model (SM) \cite{Laporta:1996mq,Moore:1984eg,Jegerlehner:2018zrj,Czarnecki:1998rc}, as well as beyond the SM (BSM) theories \cite{Lindner:2016bgg}. Furthermore, the fermion electric dipole moment (EDM) has also been analyzed in several models \cite{Pospelov:2005pr,Czarnecki:1900zz,Moyotl:2011yv,Novales-Sanchez:2016sng,Chang:2017vzi,Keus:2017ioh,Gisbert:2019ftm}. More recently,  the calculation of radiative corrections to the gluon-quark-quark $\bar{q}qg$ vertex has also become a  topic of considerable interest. Radiative corrections to the $\bar{t}tg$ coupling are expected to be considerably larger  than those of lighter quarks due to the large top quark mass \cite{Hernandez-Juarez:2018uow}. In particular  the top quark chromomagnetic dipole moment (CMDM) and  chromoelectric dipole moment (CEDM)   have been studied within the framework of the SM \cite{Martinez:2007qf},  two-Higgs doublet models (THDMs) \cite{Gaitan:2015aia}, the four-generation THDM  (4GTHDM) \cite{Hernandez-Juarez:2018uow}, models with heavy $Z$ gauge bosons \cite{Aranda:2018zis}, little Higgs models \cite{Cao:2008qd,Ding:2008nh}, the minimal supersymmetric standard model (MSSM) \cite{Aboubrahim:2015zpa}, unparticle model \cite{Martinez:2008hm}, vector like multiplet models \cite{Ibrahim:2011im}, etc.

The anomalous $\bar{q}qg$ coupling can be written as
\begin{equation}
\label{lag}
\mathcal{L}=\frac{g_s}{2} \bar{q}T^{a}\sigma _{\mu\nu} \left(  \frac{a_q}{2m_q}+i d_q \gamma ^5  \right)q G^{\mu \nu}_a,
\end{equation}
where $a_q$ and $d_q$ are the CMDM and CEDM, respectively, whereas   $G^{\mu\nu}_a$ is the gluon field strength tensor and $T^a$ are the $SU(3)$ color generators. The CMDM and CEDM are usually defined in the literature as dimensionless parameters  \cite{Khachatryan:2016xws}
\begin{align}
\hat{\mu}_q &\equiv\frac{m_q}{g_s}\widetilde{\mu}_q ,\nonumber\\ \hat{d}_q&\equiv\frac{m_q}{g_s}  \widetilde{d}_q,
\end{align}
where $\hat{\mu}_q= a_q /2$ and $\hat{d}_q=m_q d_q$.  In the case of the quark top, the most recent experimental bounds from CMS are  $-0.014<\hat{\mu}_t<0.004$  and $-0.020<\hat{d}_t<0.012 $ \cite{CMS:2018jcg}.

In the SM, the CMDM is induced at the one-loop level or higher orders via electroweak (EW) and QCD contributions. On the other hand, the CEDM is induced up to the three-loop level \cite{Czarnecki:1997bu,Khriplovich:1985jr} since  all the partial contributions exactly cancel out at the two-loop level \cite{Shabalin:1978rs}. The lowest order SM contributions  to the CMDM of the top quark have been  studied in \cite{Martinez:2007qf} and more recently  in \cite{Choudhury:2014lna,Aranda:2018zis}. However, there are some disagreement between those calculations (see section 2.1 of \cite{Choudhury:2014lna} and section 3.D of  \cite{Aranda:2018zis}). In particular,  authors of  Ref. \cite{Martinez:2007qf} only focus on the static CMDM, which they claim it receives an infrared finite QCD contribution, whereas  authors of Ref. \cite{Choudhury:2014lna}  argue that the on-shell CMDM ($q^2=0$) has no sense in perturbative QCD as this contribution diverge, so they consider  the off-shell CMDM ($q^2\neq0$). Even more,   the analytical results presented  in \cite{Martinez:2007qf}, \cite{Choudhury:2014lna} and \cite{Aranda:2018zis} disagree. 

In the experimental side, the advent of the LHC has triggered the interest on the  anomalous $\bar{t}tg$ couplings, which have become an important area of study in experimental particle physics. Searches for any deviation to the SM $\bar{t}t$ production has made it possible to set constraints on the top quark CMDM and CEDM, which  are regularly improved  \cite{Sirunyan:2019wka,Khachatryan:2016xws,CMS:2018jcg}. In fact, the current upper bounds   have been enhanced by one order of magnitude  as compared to the previous ones \cite{Khachatryan:2016xws}. Hopefully, more tight constraints on these  top quark properties, closer to the SM predictions, would be achieved in the future,  and thus a more precise and unambiguous prediction of the SM contributions to the top quark  CMDM  is mandatory. Also, since contributions to the CMDM in extension theories could give rise to a sizeable enhancement, a precise determination of such contributions is in order.

In this work we  present a new calculation of the SM one-loop contributions to the  quark CMDM, which is aimed to address some ambiguities of previous results. Our calculation is done via both a renormalizable linear $R_\xi$ gauge and the background field method (BFM), which allows one to verify that the  off-shell CMDM of a quark is gauge independent, which in turn is a necessary condition for a valid observable quantity. The rest of the  manuscript is organized as follows. In Sec. \ref{calc}, we present the main steps of the analytical calculation, stressing any disagreement with previous results. The corresponding loop functions are presented in terms of Feynman parameter integrals, Passarino-Veltman scalar functions  and closed form results in   \ref{Loopfunctions}, which may be useful for a numerical cross-check.  In Sec. \ref{SMnumerical}, we present a numerical analysis  and discussion of the behavior of the  CMDM of SM quarks, with emphasis on the top quark one. The conclusions are presented in Sec. \ref{conclusions}.

\section{Analytical results}\label{calc}

\subsection{Quark-gluon vertex function}

\begin{figure}
\center\includegraphics[width=4cm]{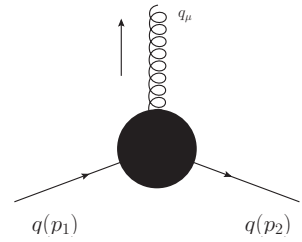}
\caption{Notation for the quark-gluon vertex. \label{quark-gluon_vertex}}
\end{figure}

\begin{figure*}[hbt!]
\begin{center}
\includegraphics[height=4cm]{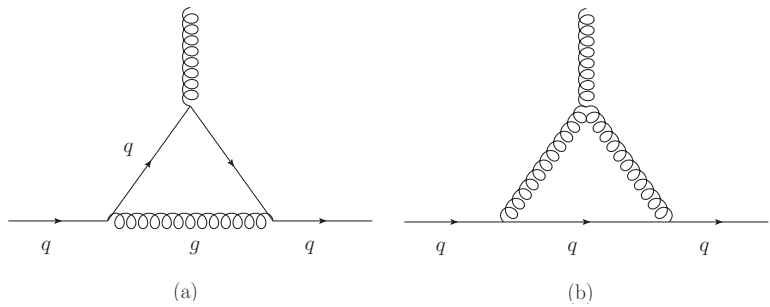}
\caption{One-loop Feynman diagrams for the QCD contributions to the CMDM of quarks: (a)QED-like diagram and (b)three-gluon diagram. For the BFM  the external gluon is replaced by its background field $g_B$.}
\label{QCDDiagrams}
\end{center}
\end{figure*}

\begin{figure*}[hbt!]
\begin{center}
\includegraphics[width=14cm]{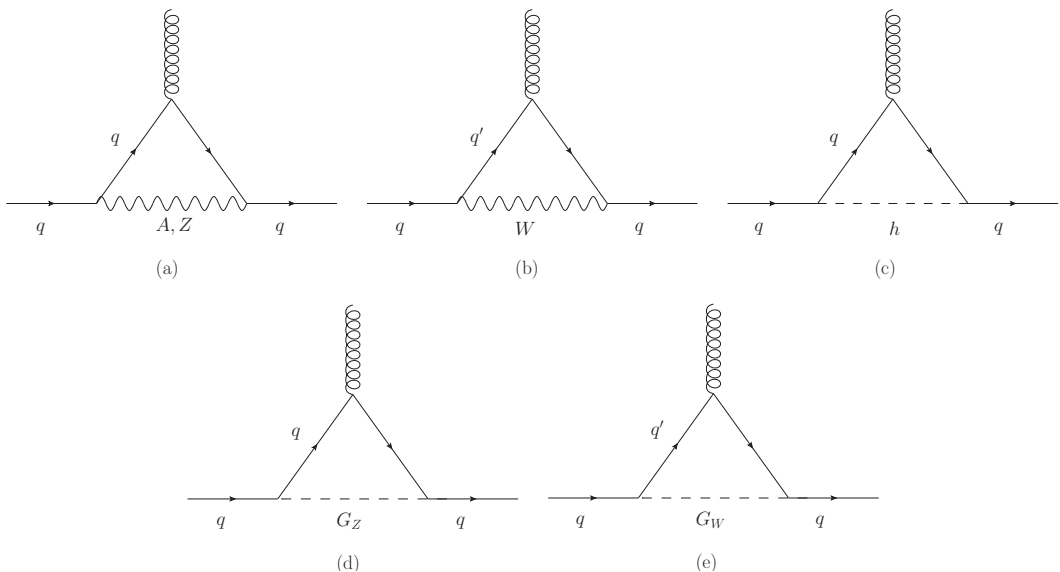}
\caption{Feynman diagrams for the SM electroweak contributions to the CMDM of quarks at the one-loop level in a renormalizable linear $R_\xi$ gauge: (a) and (b)gauge boson exchange, (d) and (e) Goldstone boson exchange, and (c)Higgs boson exchange.  For the BFM the  gluon is replaced by its background field $g_B$.
\label{EWDiagrams}}
\end{center}
\end{figure*}

For off-shell quarks and gluon, the most general $CP$ conserving  quark-gluon vertex function  can be cast as \cite{Davydychev:2000rt}
\begin{equation}
\label{funvert}
\Gamma^a_\mu=g_s T^a\Gamma_\mu=g_s T^a(\Gamma^L_\mu+\Gamma^T_\mu),
\end{equation}
where the longitudinal $\Gamma^L_\mu$ (transverse $\Gamma^T_\mu$) vertex function can be decomposed  into four (eight) independent form factors $\lambda_i$ ($\tau_i$), which depend on $q^2$, $p_1^2$ and $p_2^2$, where we follow the notation of Fig. \ref{quark-gluon_vertex}.  When $p_1^2=p_2^2$, only the following linear independent terms survive:
\begin{align}
{\Gamma^L}^\mu=&\lambda _1 \gamma ^{\mu }+\lambda _2 p^{\mu }
   \slashed{p}+\lambda _3
   p^{\mu },
\end{align}
and
\begin{align}
{\Gamma^T}^\mu=&\tau _3
   \left(q^2 \gamma ^{\mu }-q^{\mu } \slashed{q}\right)+i \tau _5
   q^{\nu } {\sigma^\mu}_\nu-i\tau _7\,
   p^{\mu } p_1^{\nu }p_2^{\lambda } \sigma_{\lambda \nu }
   \nonumber\\&+\tau _8 \left(-i  p_1^{\nu }p_2^{\lambda }
   \gamma ^{\mu }\sigma_{\nu \lambda }+p_2^{\mu }
   \slashed{p_1}-p_1^{\mu } \slashed{p_2}\right),
\end{align}
where $p=p_1+p_2$. For on-shell quarks  $\Gamma_\mu$ is enclosed by  Dirac spinors and we arrive at the standard form
\begin{align}
\label{vert}
\overline{u}(p_2)\Gamma^{\mu} u(p_1)&=
F_1(q^2)\overline{u}(p_2)\gamma^\mu u(p_1)\nonumber\\&
-i F_2(q^2)\overline{u}(p_2)\left(\sigma^{\mu\nu} q_{\nu}\right) u(p_1),
\end{align}
where the static CMDM can be obtained from the Pauli form factor as follows $a_q=-2 m_q F_2(0)$, but in this work we are interested in the case with $q^2\ne 0$.
At the one-loop level, $a_q(q^2)$ receives  QCD and EW contributions from the SM  via the Feynman diagrams  depicted in Figs. \ref{QCDDiagrams} and \ref{EWDiagrams}, respectively. We will address below the issue of the gauge independence of $a_q(q^2)$, which is necessary to  have a valid observable quantity.

\subsection{The off-shell CMDM of quarks}
While on-shell Green functions are  gauge invariant and gauge independent, this is not necessarily  true for off-shell Green functions as they do not correspond to a physical process but just to an amputated set of Feynman diagrams. The first systematic approach to obtain well-behaved off-shell Green functions out of which valid observable quantities can be extracted was the so-called pinch technique (PT) \cite{Cornwall:1981zr,Cornwall:1989gv,Binosi:2009qm}, which is based  on a diagrammatic method that combines  self-energy, vertex and box diagrams associated to a physical process to remove any gauge dependent term. It was later shown that  at the one-loop level the results obtained via the PT coincide with those obtained by the BFM via the Feynman-'t Hooft gauge ($\xi_Q=1$) \cite{Denner:1994xt,Pilaftsis:1996fh}.  In this work we use the later approach to obtain a gauge independent quark CMDM as it is simpler in computational grounds.

Since we are interested in the CMDM of quarks for an off-shell gluon, we need to verify  that the contributions of Figs. \ref{QCDDiagrams} and \ref{EWDiagrams} are  gauge independent and thus  provide an observable quantity. With this aim, to cross-check our result, for our calculation we use both the conventional renormalizable linear $R_\xi$ gauge and the BFM \cite{Denner:1994xt} with arbitrary gauge parameters.  Indeed, both computations are technically identical except for  the contribution of the three-gluon diagram of Fig. \ref{QCDDiagrams} as for the remaining diagrams the only dependence on the  gauge parameters arises from the gauge boson propagators.  The main outline of our calculation is as follows. We first wrote out the amplitude for each contributing Feynman diagram for arbitrary $\xi$ ($\xi_Q$)  and then used the  Passarino-Veltman reduction scheme to perform the integration over the four-momen\-tum space. Once the gauge parameter independence of $\hat \mu_q(q^2)$ was verified, we used Feynman parameter integration in the unitary gauge to obtain an alternate result, which can be used for  a numerical cross-check. The loop functions are thus presented  in terms of Feynman parameter integrals and  Pa\-ssa\-rino-Veltman scalar functions. For the latter we include the results in terms of closed form functions if possible. The Dirac algebra and tensor reduction was done with the help of  FeynCalc \cite{Shtabovenko:2016sxi} and Package-X \cite{Patel:2015tea}, which prove very helpful to verify explicitly that the  gauge parameter drops out from the calculation.

Below,  $p_1$ ($p_2$) denotes the four-momentum of the ingoing (outgoing) quark, whereas $q$ is the gluon four-momentum as shown in Fig. \ref{quark-gluon_vertex}. It is understood that in  the $D$-dimensional integrals, the volume element $d^D  k$ is accompanied by a factor of $\mu^{4-D}$,  with $\mu$ the scale of dimensional regularization, that drops out from the final results as they are ultraviolet finite. Also,  a small imaginary part $i\epsilon$ must be added to the propagators.  Below, the gauge parameters are  indistinctly denoted by $\xi$ as they cancel out for each partial contribution.

\subsection{QCD contribution}\label{QCDcalculations}

The contribution to the vertex function $\Gamma^\mu$ from diagram \ref{QCDDiagrams}(a) is
\begin{align}
   \Gamma^\mu_{{\rm QCD}_1}&=  \frac{ig_s^2}{6} \int \frac{d^D  k}{(2\pi)^D}\frac{\gamma_\beta\left( \slashed{q_2} +m_q\right) \gamma ^\mu \left( \slashed{q_1} +m_q\right)\gamma_\alpha}{\left(q_2^2 -m_q^2\right)\left(q_1^2 -m_q^2\right)\left(k^2\right)}\nonumber\\&
 \times P^{\alpha\beta}(k),
\end{align}
where
$P^{\lambda\rho}(p)=
\left(g^{\lambda\rho}+(\xi-1)\frac{p^\lambda p^{\rho}}{p^2}\right)$,
 $q_i=k+p_i$, and  $m_q$ is the quark mass.
In this case we have three color generators $T^a$, which simplifies as follows
\begin{equation}
\label{color1}
    T ^b_{jn}T^a_{nm}T^b_{mi}= \frac{1}{2}\delta_{ji}T^a_{nn}-\frac{1}{2N}T^a_{ji}=-\frac{1}{6} T^a_{ji},
\end{equation}
where we used $T ^b_{jn}T^b_{mi}=\frac{1}{2}\left( \delta_{ji}\delta_{nm}-\frac{1}{N}\delta_{jn}\delta_{mi}
 \right)$ and ${\rm Tr}\left[ T^a\right]=0$, whereas $N$ stands for the quark color number.

After four-momentum integration, the $\xi$ parameter drops out and the following contribution to the CMDM is obtained:
\begin{equation}
\label{QCD1at}
a^{{\rm QCD}_1}_{q}\left(q^2\right)=\frac{\text{$\alpha _s$}}{6 \pi}\mathcal{F}^{{\rm QCD}_1}_{q}\left(q^2\right),
\end{equation}
where the function $\mathcal{F}^{{\rm QCD}_1}_{q}$ is presented in  \ref{Loopfunctions}.\footnote{From now on, all the corresponding loop functions appearing in the contributions to $a_q\left(q^2\right)$, denoted by calligraphy letters, will be presented in terms of Feynman parameter integrals, Passarino-Veltman scalar functions and closed form results in appendices \ref{FeynIntegrals}, \ref{PV}, and \ref{analy}, respectively, including the results for $q^2=0$.} When $q^2=0$, it is straightforward to obtain

\begin{equation}
a^{{\rm QCD}_1}_q(0)=-\frac{\text{$\alpha _s$} }{12 \pi},
\end{equation}
which agrees with the well-known QED result  after the replacement of  the electric charge $e$ by the strong coupling constant $\alpha_s$ and the insertion of the color factor of Eq. \eqref{color1}.

The non-abelian contribution to the quark CMDM from diagram \ref{QCDDiagrams}(a) can be obtained from the following vertex function:

\begin{align}
\Gamma^\mu_{{\rm QCD}_2}&=  -\frac{i3g_s^2}{2} \int \frac{d^D k}{(2\pi)^D} \frac{\gamma_\rho \left(-\slashed{k}+m_q\right)\gamma_\lambda\Sigma^{\lambda\rho\mu}}{\left(q_2^2 \right) \left(k^2-m_q^2 \right) \left(q_1^2 \right)},
\end{align}
with
\begin{align}
\Sigma^{\lambda\rho\mu}&=\Big(
g^{\alpha \mu }
\left(p_2-k-2p_1\right)^{\beta
   }+g^{\beta \mu }
   \left(p_1-k-2p_2\right)^{\alpha }
   \nonumber\\&+g^{\alpha
   \beta } \left(2k+p_1+p_2\right)^{\mu }+\frac{1}{\xi
   }\left(g^{\alpha \mu } q_2^{\beta }+g^{\beta \mu } q_1^\alpha\right)\Big)\nonumber\\&\times P_\alpha^{\lambda}(q_1)P_\beta^{\rho}(q_2),
\end{align}
for the BFM, whereas for the linear $R_\xi$ gauge we must drop the $\frac{1}{\xi}$ term between the parenthesis.

As for the corresponding  color factor, it was worked out as follows
\begin{equation}
T^m_{jb}T^n_{bi}f^{anm}=-2i\text{Tr}\left[ T^m\left[T^n,T^a\right]\right] T^m_{jb}T^n_{bi}  = -\frac{i3}{2}T^a_{ji}.
\end{equation}
After four-momentum integration, our result for the quark CMDM is given as
\begin{equation}
\label{QCD2at}
a^{{\rm QCD}_2}_q\left(q^2\right)=\frac{3\alpha _s}{2 \pi}\mathcal{F}^{{\rm QCD}_2}_{q}\left(q^2\right),
\end{equation}
which disagrees with the result obtained in  \cite{Choudhury:2014lna} as there is a disagreement with the color factor used by those authors.

When $q^2=0$,
Eq. \eqref{QCD2at} yields an infrared divergent result for $q^2=0$:

\begin{align}
\mathcal{F}_q^{{\rm QCD}_2}(0)&= \frac{1}{2} \left(\frac{1}{\epsilon }+\log \left(\frac{\mu
   ^2}{m_q^2}\right)+3\right),
\end{align}
where $\epsilon$ is the pole of dimensional regularization. Therefore, the contribution of the three-gluon diagram is not well defined when $q^2=0$ as it was also pointed out in \cite{Choudhury:2014lna}.  Since QCD contributions to the CMDM are proportional to the strong running coupling constant $\alpha_s\left(q^2\right)$, such contributions have not perturbative sense at $q^2=0$ but at a scale where a perturbative calculation is valid. 

\subsection{Electroweak contribution}
We now present the calculation of the contributions to the quark CMDM  induced through the Feynman diagrams of Fig. \ref{EWDiagrams}.  We note that  the  diagrams with  photon,  $Z$ gauge boson, $\varphi_Z$ Goldstone boson, and Higgs boson exchange are similar to those inducing  a lepton anomalous MDM \cite{Moore:1984eg,Hollik:1998vz},  but with the external photon replaced by a gluon. Therefore, the CMDM   just differ by the coupling constants $g_s$  and the $SU(3)$ generators $T^a_{ij}$ instead of the electric charge. Thus our result must reproduce that of the anomalous MDM of a lepton.  As far as the diagrams with  $W^\pm$ gauge boson and $G_W$ Goldstone boson exchange are concerned, the lepton anomalous MDM has no analogous contributions.
\subsubsection{Photon exchange}
The corresponding contribution to the $\bar{q}q g$ vertex function can be written as

\begin{align}
\Gamma_A^\mu&=  -i e^2 Q_q^2 \int \frac{d^D k}{(2\pi)^D} \frac{\gamma_\beta\left( \slashed{q_2} +m_q\right) \gamma ^\mu \left( \slashed{q_1} +m_q\right)\gamma_\alpha}{\left(q_2^2 -m_q^2\right)\left(q_1^2 -m_q^2\right)\left(k^2\right)}\nonumber\\&
 \times  \left(g^{\alpha\beta}+(\xi-1)\frac{k^\alpha k^\beta}{k^2}\right),
\end{align}
with $Q_q$ the quark electric charge in units of $e$. After a straightforward calculation, we arrive at a gauge-parameter independent result for an on-shell gluon. It
reads
\begin{equation}
\label{Photonexchangeat}
a^A_q\left(q^2\right)=-\frac{e^2 Q_q^2}{4 \pi ^2 }\mathcal{F}_{q}^A\left(q^2\right),
\end{equation}
which gives  a result similar to that of the lepton anomalous MDM for $q^2=0$:
\begin{equation}
a^A_q(0)=\frac{\alpha Q_q^2}{2\pi}.
\end{equation}
We note that the electric charge factor $Q_q^2$ is missing  in the corresponding  result of \cite{Choudhury:2014lna}. However, since the internal photon of diagram \ref{EWDiagrams}(a) is attached to two quark lines, such a factor must appear in this contribution.

\subsubsection{$Z$ gauge boson and $\varphi_Z$ Goldstone boson exchange}
We now present the calculation for  the contributions of the loops with the neutral $Z$ gauge boson and its associated Goldstone boson $\varphi_Z$ [diagrams (a) and (d) of Fig. \ref{EWDiagrams}]. The corresponding contributions need to be added up to cancel out the dependence on the gauge parameter $\xi$.
As far as the diagram with $Z$ gauge boson exchange is concerned,  the $\bar{q}qg$ vertex function in terms of the axial (vector) $g_A^q$ $(g_V^q)$ couplings reads:
\begin{equation}
\label{Zvertex}
  \Gamma^\mu_Z=  \frac{-i g^2}{c_W^2} \int \frac{d^D k}{(2\pi)^D} \frac{\Xi^\mu}{\left(q_2^2 -m_q^2\right
  )\left(q_1^2 -m_q^2\right)\left(k^2 - m_Z ^2\right)},
\end{equation}
where
\begin{align}
\Xi^\mu&=\gamma^\beta\left(g_V^q-g_A^q\gamma^5\right)\left( \slashed{q_2} +m_q\right) \gamma ^\mu \left( \slashed{q_1} +m_q\right)\gamma^\lambda\nonumber\\&\times \left(g_V^q-g_A^q\gamma^5\right)
\left(g_{\beta\lambda}+(\xi-1)\frac{k_\beta k_\lambda}{k^2-\xi m_Z^2}\right),
\end{align}
the vector and axial vector couplings are defined as
\begin{equation}
g_V^q= \frac{1}{2}T_{3_q} -Q_{q} s_W^2  \text{,}\quad g_A^q=\frac{1}{2}T_{3_q},
\end{equation}
with $T_{3_q}$ the weak isospin ($T_{3_u}=\frac{1}{2}$, $T_{3_d}=-\frac{1}{2}$).
On the other hand, the contribution from the diagram with $\varphi_Z$ Goldstone boson exchange is
\begin{align}
\label{GZvertex}
  \Gamma^\mu_{\varphi_Z}&=  \frac{-i g^2 m_q^2}{4m_W^2} \int \frac{d^D k}{(2\pi)^D}\gamma^5\left( \slashed{q_2} +m_q\right) \gamma ^\mu \left( \slashed{q_1} +m_q\right)\gamma^5\nonumber\\
  &\times
  \frac{1}{\left(q_2^2 -m_q^2\right
  )\left(q_1^2 -m_q^2\right)\left(k^2 - \xi m_Z ^2\right)}.
\end{align}
The explicit integration in the four-momentum space shows that the dependence on the $\xi$ gauge parameter cancels out  after adding up the contributions of the $Z$ and $\varphi_Z$ exchange diagrams. The total contribution is thus given by
\begin{align}
\label{Zexchangeat}
a_q^Z\left(q^2\right)&=\frac{\sqrt{2}  G_F m_q^2 }{\pi ^2}\Bigg(\left(g^q_A\right)^2\mathcal{A}_q^Z\left(q^2\right)+\left(g^q_V\right)^2\mathcal{V}_q^Z\left(q^2\right)\Bigg),
\end{align}
whereas the result  for $q^2=0$  is analogue to the $Z$ contribution to the anomalous MDM of a lepton \cite{Moore:1984eg}.
There is agreement with the  calculation presented in \cite{Aranda:2018zis}, but there is no agreement with the result of \cite{Choudhury:2014lna} as those authors use the Feynman-'t Hooft gauge propagator for the $Z$ gauge boson but seem to omit  the $\varphi_Z$ Goldstone boson exchange contribution.

\subsubsection{$W^\pm$ gauge boson and $\varphi^\pm$ Goldstone boson exchange}
We now calculate the contribution from the Feynman diagrams (b) and (e) of Fig. \ref{EWDiagrams} as both contributions must be added up in order to drop out the dependence on the gauge parameter $\xi$. For the diagram with $W$ gauge boson exchange, the $\bar{q}qg$ vertex function can be written as
\begin{align}
\label{VertW}
  \Gamma^\mu_W&=  \sum_{q'}\frac{-i g^2 \left| V_{qq'}\right| {}^2}{2} \int \frac{d^D k}{(2\pi)^D}\Pi^\mu\nonumber\\&\times \frac{1}{\left(q_2^2 -m_{q'}^2\right)\left(q_1^2 -m_{q'}^2\right)\left(k^2 -m_W ^2\right)},
\end{align}
with
\begin{align}
\Pi^\mu&=\gamma^\beta P_L\left( \slashed{q_2} +m_{q'}\right) \gamma ^\mu \left( \slashed{q_1} +m_{q'}\right)\gamma^\lambda P_L\nonumber\\&\times
\left(g_{\beta\lambda}+(\xi-1)\frac{k_\beta k_\lambda}{k^2-\xi m_W^2}\right),
\end{align}
whereas the contribution of the $\varphi^\pm$ Goldstone boson reads
\begin{align}
\label{VertGW}
  \Gamma^\mu_{\varphi^\pm}&=  \sum_{q'}\frac{i g^2 \left| V_{qq'}\right| {}^2}{2m_W} \int \frac{d^D k}{(2\pi)^D}\Pi^{'\mu}\nonumber\\&\times \frac{1}{\left(q_2^2 -m_{q'}^2\right)\left(q_1^2 -m_{q'}^2\right)\left(k^2 -\xi m_W ^2\right)},
\end{align}
with
\begin{align}
\Pi^{'\mu}&=\left(m_q P_L-m_{q'}P_R\right)\left( \slashed{q_2} +m_{q'}\right) \gamma ^\mu \left( \slashed{q_1} +m_{q'}\right)\nonumber\\
&\times\left(m_q P_R-m_{q'}P_L\right),
\end{align}
where $P_{L,R}$ is the chirality projector, $q'$ stands for the internal quark and $V_{qq'}$ is the CKM matrix element.

Again after four-momentum integration, the  gauge parameter drops out and  the following gauge independent contribution to the  quark CMDM is obtained:
\begin{equation}
\label{Wexchangeat}
a_q^W\left(q^2\right)=\sum_{q'} \frac{G_F m_{q}^2  \left| V_{qq'}\right| {}^2}{4 \sqrt{2} \pi ^2 }\mathcal{F}_{qq'}^W\left(q^2\right),
\end{equation}
with the dominant term arising from the diagonal CKM matrix element ($V_{qq} \approx1$). There is no agreement with the result of \cite{Choudhury:2014lna} as those authors  consider that the external and internal quark masses are  degenerate.

\subsubsection{Higgs boson exchange}
The remaining SM contribution to the quark CMDM arises from the diagram with Higgs boson exchange, which is gauge independent. The corresponding contribution to the $\bar{q}qg$ vertex function is given by

\begin{equation}
  \Gamma^\mu_h=  \frac{i g^2 m_q^2}{4m^2_{W}} \int \frac{d^D k}{(2\pi)^D} \frac{\left( \slashed{q_2} +m_q\right) \gamma ^\mu \left( \slashed{q_1} +m_q\right)}{\left(q_2^2 -m_q^2\right)\left(q_1^2 -m_q^2\right)\left(k^2 -m_h ^2\right)}.
\end{equation}
The algebra is straightforward  and we obtain after four-momentum integration:
\begin{equation}
\label{Higgsexchangeat}
a^h_q\left(q^2\right)=-\frac{G_F m_q^2}{4 \sqrt{2} \pi ^2 }\mathcal{F}_q^h\left(q^2\right),
\end{equation}
which for $q^2=0$ agrees with the results presented in  \cite{Choudhury:2014lna}.

\section{Numerical evaluation and discussion}\label{SMnumerical}
We now turn to the numerical evaluation of the one-loop contribution to the CMDM  of quarks. We first present a numerical estimate of the quark CMDM in the SM, which is aimed to make a comparison with previous results, which can be useful to  settle any ambiguity.

\subsection{Off-shell CMDM of quarks in the SM}
\begin{figure*}[hbt!]
\begin{center}
\includegraphics[width=14cm]{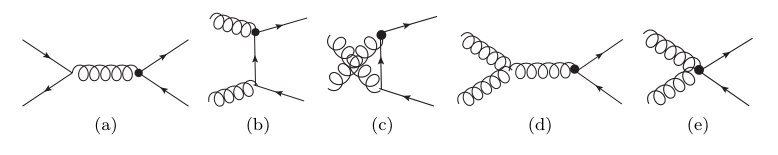}
\caption{Feynman diagrams for $\overline{t}t$ production via the Lagrangian of Fig. \ref{lag}. The dots correspond to the contributions of the top quark CMDM to the $\overline{t}t$ production}.\label{ttProductionDiagrams}
\end{center}
\end{figure*}

We first analyze the behavior of the parameter $\hat{\mu}_=a_q/2$ in the SM, which is the one studied by the experimental collaborations \cite{Khachatryan:2016xws}. Although the top quark CMDM is the one mainly studied in the literature,  for the sake of completeness we include in our analysis an estimate for all the SM quarks. Since the results for the QCD contribution have not sense in perturbative calculations at $q^2=0$, as pointed out above, we study the  case $q^2\neq 0$. Anomalous top quark couplings have been studied at the LHC through  $\overline{t}t$ production \cite{Hioki:2009hm,Kamenik:2011dk,Bernreuther:2013aga,CMS:2018jcg}, moreover, its effects to the $\overline{t}t$ cross section have been analyzed in \cite{Cheung:1995nt,Haberl:1995ek}. The transition CMDM could contribute at the leading order through the diagrams of Fig. \ref{ttProductionDiagrams}, where the top quark CMDM contributions are marked by a dot and we include the $gg\overline{t}t$ interaction arising from Lagrangian \eqref{lag}.  Since the outgoing top quarks  are on-shell,  the gluon four-momentum  in the $s$-channel diagrams obeys  $\|q\|\ge 2m_t$, whereas in the $t$ and $u$ channels there are no such kinematical constraint.

We have implemented the strong coupling constant $\alpha_s \left(q^2\right)$ as the three loop approximate solution of the renormalization group equation of QCD \cite{Larin:1993tp,Prosperi:2006hx}. We consider gluon four-momentum transfer in the  30-1000 GeV region, where $\alpha_s\sim 0.1$, since for $\|q\|$ less than around 1 GeV the theory becomes strongly interacting \cite{Tanabashi:2018oca}. In addition, at next-to-leading order QCD calculations, EW corrections are neglected, so only  the pure QCD contribution to the CMDM of quarks would be relevant.

For the numerical analysis we use the results in terms of Passarino-Veltman scalar functions, which were evaluated via the LoopTools \cite{Hahn:1998yk} and Collier \cite{Denner:2016kdg} packages, though we cross-check with the results obtained by numerical integration of the Feynman parameter results, which however shows more numerical instability.

\subsubsection{Light quarks CMDM}

\begin{figure*}[hbt!]
\begin{center}
{\includegraphics[width=.9 \textwidth]{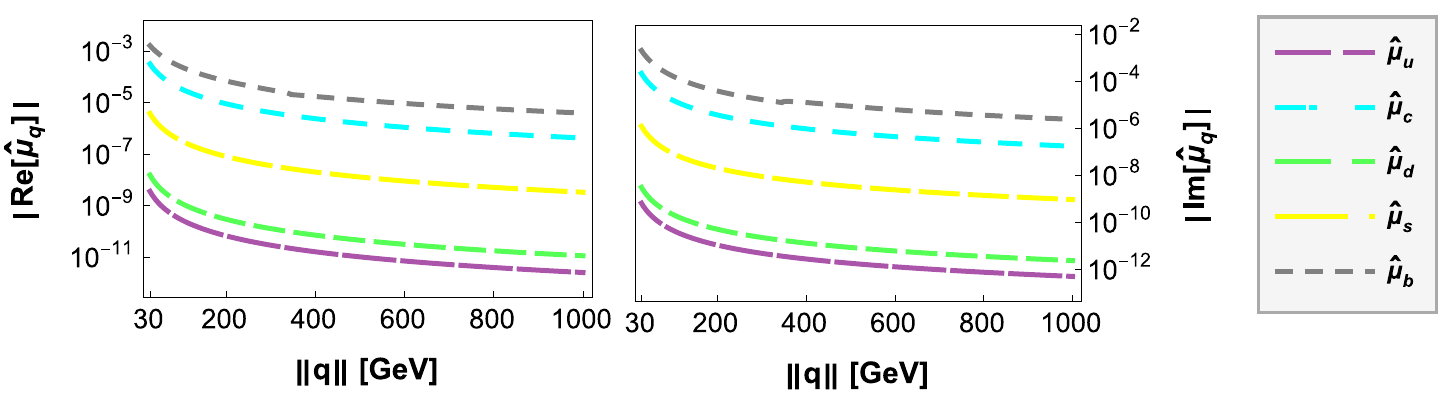}}
\caption{Real (left plot) and imaginary (right plot) parts of the light quarks CMDM $\widehat{\mu}_q$ as function of the transfer momentum  of the gluon. }
\label{LightQuarksPlot}
\end{center}
\end{figure*}

We show in Fig. \ref{LightQuarksPlot} the behavior of the real $\text{Re}\left[ \widehat{\mu}_q \right]$  and imaginary $\text{Im}\left[ \widehat{\mu}_q \right]$  parts of the CMDM of the light SM quarks as functions of the gluon transfer momentum  $\|q\|$. We observe that in both cases the largest estimates correspond to the $b$ quark CMDM, whereas the smallest estimates are obtained for the $u$ and $d$ quarks. This stems from the fact that  the CMDM is proportional to the quark mass for $q^2\neq0$. We also note that the real and imaginary parts of $\hat \mu_q$ are about the same magnitude for all the light quarks. Numerical predictions for the CMDM of light quarks  are shown in Table \ref{CMDM} for some selected values of $\|q\|$.

\begin{figure*}[hbt!]
\begin{center}
{\includegraphics[width=.9 \textwidth]{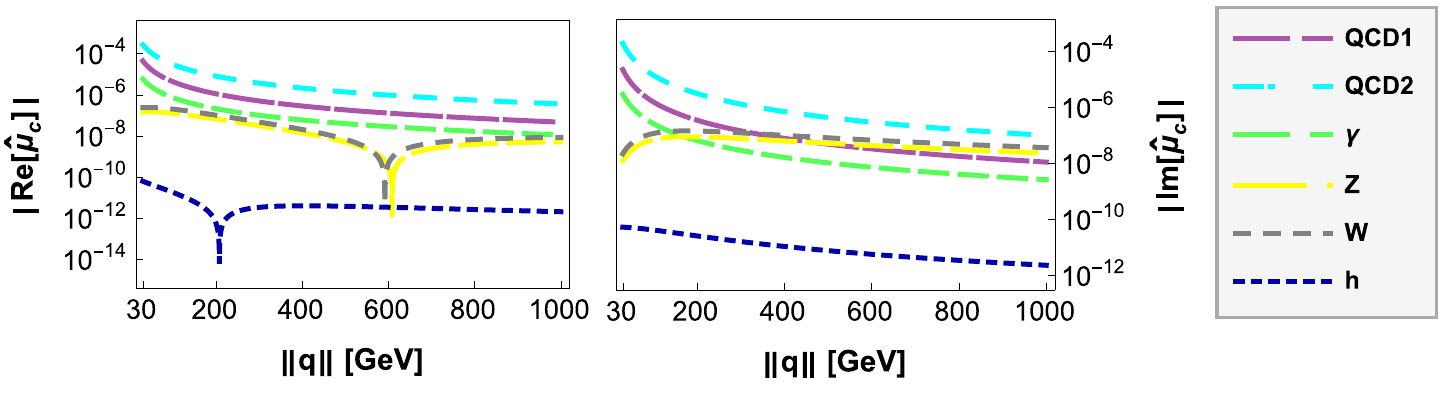}}
\caption{Real (left plot) and imaginary (right plot) parts of the SM one-loop partial contributions to $\hat\mu_c$  as functions of the transfer momentum of the gluon $\|q\|$.}
\label{LightQuarksIndividual}
\end{center}
\end{figure*}

We now turn to analyze the behavior of the partial contributions to $\hat\mu_q$ for a light quark. Thus, by way of illustration, we show in Fig. \ref{LightQuarksIndividual} the real  and imaginary  parts of the partial contributions  to the $c$ quark CMDM. All other light quark's contributions exhibit a similar behavior, though there are slight changes for the $b$ quark as explained below. We first note that  the dominant contributions arise  from the triple gluon vertex (the so-called ${\rm QCD}_2$ contribution), though at high energies the ${\rm QCD}_1$, $\gamma$, $Z$ and $W$ contributions are of similar size. In particular, the imaginary parts of the EW gauge bosons contributions are slightly larger than the one of the ${\rm QCD}_1$ contribution for $\|q\|\gtrsim 600$ GeV, whereas the real parts of both QCD contributions  dominate in all the studied energy interval.  On the other hand, the Higgs boson contributions are the smallest ones: for  the $u$ and $d$ quarks, such contributions  are negligibly small,  of the order of $10^{-20}-10^{-21}$. Note that for $\|q\|>30$ GeV all the partial contributions to $\hat\mu_q$ develop an imaginary part as $q^2>4m^2$, with $m$  the mass of the virtual particles attached to the external gluon, except for the $W$ contribution to $\hat\mu_b$, which is purely real for $q^2<4m_t^2$ as long as one neglects the contributions of the loops with internal  $u$ and $c$ quarks.

\subsubsection{Top quark CMDM}

\begin{figure*}[hbt!]
\begin{center}
{\includegraphics[width=.9 \textwidth]{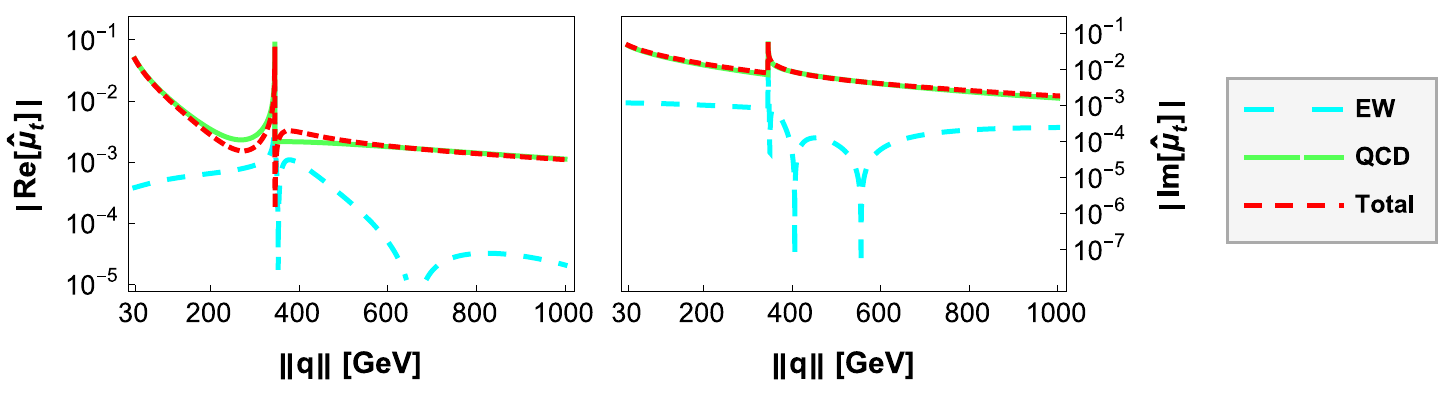}}
\caption{Real (left plot) and imaginary (right plot) parts of the EW, QCD and total contributions to the top quark CMDM $\widehat{\mu}_t$ as function of the transfer momentum norm of the gluon $\|q\|$. }
\label{TopCMDM}
\end{center}
\end{figure*}

We now turn to analyze the behavior of the CMDM of the top quark. We first show in Fig. \ref{TopCMDM}   $\hat \mu_t$ as a function of $\|q\|$ as well as its partial QCD and EW contributions. We observe that both the real and imaginary parts  are  dominated by the QCD contributions, though the real part of the EW contribution is  of comparable size around the threshold $\|q\|=2m_t$, where all the contributions show a peak due to  a flip of sign.  Both the QCD and EW contributions decrease as $\|q\|$ increases: above the $2m_t$ threshold the real part of the EW contribution becomes negligible, whereas its imaginary part is  about one order of magnitude below the imaginary part of the QCD contribution for $\|q\|\sim 1000$ GeV. However, at very large $\|q\|$ (  much larger than $1000$ GeV) the imaginary part of the EW contribution becomes dominant since the imaginary part of the QCD contribution decreases quickly at very high energies.

\begin{figure*}[hbt!]
\begin{center}
{\includegraphics[width=.9 \textwidth]{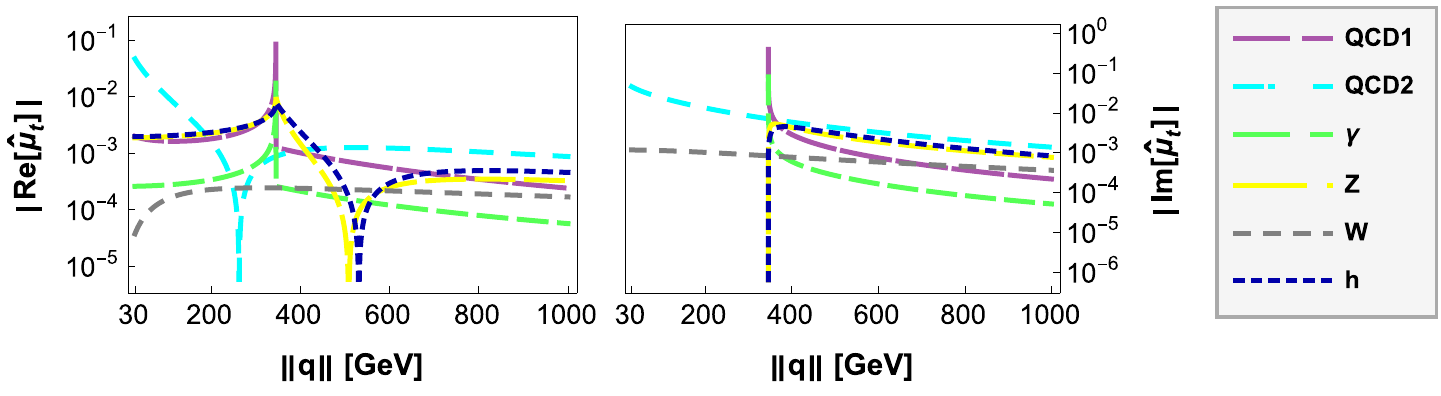}}
\caption{Real (left plot) and imaginary (right plot) parts of the SM one-loop partial contributions to the top quark CMDM $\widehat{\mu}_t$ as functions of the transfer momentum  of the gluon $\|q\|$.}
\label{TopIndividual}
\end{center}
\end{figure*}

We now show in Fig. \ref{TopIndividual}  the real  and imaginary parts  of  all the partial contributions to $\hat\mu_t$ as functions of $\|q\|$. As far as the real parts are concerned, we observe that at low and high energies the ${\rm QCD}_2$ contribution dominates, but around $\|q\|=2m_t$ the $Z$ and $h$ contributions become the dominant ones, which explains the behavior of the  EW contribution shown in Fig. \ref{TopCMDM} at $\|q\|\simeq 2m_t$. Nevertheless such contributions are of opposite sign and they tend to cancel each other out. On the other hand, as for the imaginary contributions, below the threshold $\|q\|=2m_t$ all but the ${\rm QCD}_2$ and $W$ contributions vanish and above this threshold the $Z$ and $h$ contributions develop imaginary parts  of the same order of magnitude than that of  the three-gluon contribution (${\rm QCD}_2$), which remains slightly larger as $\|q\|$ increases. We can  conclude that the QCD contributions is always  dominant,  nevertheless  the imaginary part of the EW contribution become comparable to the QCD one at high energies. After the threshold $\|q\|=2m_t$ the top quark CMDM exhibits a peak due to a flip of  sign. Such a  behavior is not observed however in the CEDM of light quarks as we are studying energies  far from the threshold region.

Finally, we show in Table \ref{CMDM} the numerical estimates of $\hat\mu_q$ for all the SM quarks at a few selected values of the gluon transfer momentum $\|q\|$. As expected, the largest estimate corresponds the top quark CMDM, though the bottom and charm quarks CMDM could also be non-negligible in some energy regions.  The CMDM of all quarks is in general complex, with real and imaginary parts of comparable size, though the real parts are always slightly larger.  We have compared our numerical results  with those reported  in \cite{Aranda:2018zis} for the top quark CMDM at $q^2=\pm m_Z^2$ and find a good agreement. In this case the imaginary part of $\hat\mu(q^2)$ arises from the  ${\rm QCD}_2$ and $W$ exchange contributions, whereas the remaining contributions are purely real as  $q^2$ is below the kinematic threshold where an imaginary part is developed.

\begin{table*}[hbt!]
  \centering
  \caption{Estimates for the SM contribution to  the CMDM $\hat\mu_q$ of the  SM quarks for select values of the gluon transfer momentum $\|q\|$.}\label{CMDM}
 \begin{tabular}{cccc}
  \\\hline\hline Quark & $\|q\|=30$ GeV & $\|q\|=m_Z$ & $\|q\|=500$ GeV
  \\\hline\hline $d$ &$1.47\times10^{-8} -i 2.96 \times10^{-9}$&$1.44\times10^{-9} -i2.55  \times10^{-10}$&$4.33\times10^{-11}- i 8.21  \times10^{-12}$
  \\\hline $u$ & 3.47$\times10^{-9}-i 6.33 \times10^{-10}$ &3.35$\times10^{-10}-i5.47\times10^{-11}$& 9.94$\times10^{-12}-i$1.75$\times10^{-12}$\\\hline $s$ &$3.63\times10^{-6}- i 1.17 \times10^{-6}$&$3.8\times10^{-7}- i1.01  \times10^{-7}$&$1.23\times10^{-8}- i3.25  \times10^{-9}$
  \\\hline $c$  &$3.08\times10^{-4}- i2.10  \times10^{-4}$&$3.96\times10^{-5}- i1.87  \times10^{-5}$&$1.51\times10^{-6}- i6.07  \times10^{-7}$
  \\\hline $b$  &$1.55\times10^{-3}- i1.95  \times10^{-3}$&$2.72\times10^{-4}- i1.96  \times10^{-4}$&$1.24\times10^{-5}- i6.56  \times10^{-6}$
  \\\hline $t$  &$-4.81\times10^{-2}- i4.69  \times10^{-2}$&$-1.33\times10^{-2}- i2.66  \times10^{-2}$&$2.24\times10^{-3}- i5.43  \times10^{-3}$
   \\\hline\hline
  \end{tabular}
  \end{table*}

\section{Conclusions}
\label{conclusions}

In this work we present a new evaluation of the SM prediction of the CMDM $\hat\mu_q$ of quarks at the one-loop level, which is aimed to address some inconsistencies appearing in previous calculations.  We considered the most general case with non-zero transfer momentum of the gluon $q^2$ and the calculation was performed within both a  renormalizable linear $R_\xi$ gauge and the BFM for arbitrary gauge parameters. It was found that the off-shell CMDM is gauge independent, which assures us that it is an observable quantity. For completeness the loop integrals are presented in terms of Feynman parameter integrals, Passarino-Veltman scalar functions and closed form functions, which are useful to make  a cross-check of the numerical results. It is found that the QCD contribution arising from the Feynman diagram with a three-gluon vertex has an infrared divergence and it thus not defined at $q^2=0$, which is due to the fact that  the static CMDM has not perturbative sense, as it has also been pointed out by the authors of Refs. \cite{Choudhury:2014lna,Aranda:2018zis}. We then perform a numerical analysis and examine the behavior of the CMDM of all the SM quarks in the  region 30 GeV$<\|q\|<$ 1000 GeV, where the QCD coupling constant $\alpha(q^2)$ is of the order of $10^{-1}$. In this energy region the CMDMs  are complex in general,   with the imaginary parts being about the same order of magnitude than the real parts. Furthermore, the QCD contributions dominate over the EW contributions, which suggests that two-loop contributions can be relevant. On the other hand, the imaginary part of the EW contribution is only comparable to the QCD contribution at very high energies. Since the CMDM is proportional to the quark mass, the largest contributions correspond to the top quark CMDM, which is of the order of $10^{-2}-10^{-3}$, with the imaginary part of the EW contributions of the same size than the QCD contributions around the threshold $q^2=4m_t^2$.

{\bf Note Added:}
After this work was submitted, we became aware that a very related paper \cite{Aranda:2020tox}, that focuses on the top quark CMDM only, had been recently posted to the preprint archive. We have found that our analytical results and numerical estimates for the top quark CMDM agree with those presented in that paper, which only considers the case with $q^2=\mp m_Z^2$.

\begin{acknowledgements}
We acknowledge support from Consejo Nacional de Ciencia y Tecnolog\'ia and Sistema Nacional de Investigadores. Partial support from Vicerrector\'ia de Investigaci\'on y Estudios de Posgrado de la Ben\'emerita Universidad Aut\'onoma de Puebla is also acknowledged.
\end{acknowledgements}

\onecolumn
\appendix
\section{Analytic results for the loop functions}
\label{Loopfunctions}
We now present the results for the loop functions appearing in the contributions to the CMDM of quarks discussed in Sec. \ref{calc} in term of Feynman parameter integrals, Passarino-Veltman scalar functions, and closed form functions. For sake of completeness we also include the results for $q^2=0$.

\subsection{Feynman parameter integrals}\label{FeynIntegrals}
We note that this calculation was done via the unitary gauge as the result are gauge independent, which was explicitly verified via the Passarino-Veltman reduction scheme. Therefore, in the EW sector we only computed by this method the Feynman diagrams (a) through (c) of Fig. \ref{EWDiagrams}.

We introduce the definition $r_{a,b}=m_a/m_b$ and present the loop functions for the QCD contributions to the CMDM of quarks. Feynman diagram \ref{QCDDiagrams}(a) yields the loop function [Eq. \eqref{QCD1at}]:

\begin{equation}
\mathcal{F}^{{\rm QCD}_1}_{q}\left(q^2\right)=m_q^2\int_0^1\int_0^{1-u}\frac{ (u-1) u}{  m_q^2 (u-1)^2+q^2 v
   (u+v-1)}dvdu,
\end{equation}
whereas  Feynman diagram \ref{QCDDiagrams}(b) gives [Eq. \eqref{QCD2at}]:
\begin{equation}
\mathcal{F}^{{\rm QCD}_2}_{q}\left(q^2\right)=m_q^2\int_0^1\int_0^{1-u} \frac{(u-1)u}{  m_q^2 u^2 - q^2 v (1- u - v)}dvdu,
\end{equation}
which for $q^2=0$ reduces to
\begin{equation}
\mathcal{F}^{{\rm QCD}_2}_q(0)=\int^1_0 \frac{  (1-u)^2 }{  u}du.
\end{equation}
As far as the EW contributions to the CMDM of quarks are concerned, the Feynman diagram with photon exchange of Fig. \ref{EWDiagrams}(a) gives [Eq. \eqref{Photonexchangeat}]:
\begin{equation}
\mathcal{F}_{q}^A\left(q^2\right)=m_q^2\int_0^1\int_0^{1-u}\frac{ (u-1) u}{m_q^2 (u-1)^2+q^2 v (u+v-1)}dvdu,
\end{equation}
whereas the $Z$ boson exchange diagram gives [Eq. \eqref{Zexchangeat}]
\begin{equation}
\mathcal{A}_q^Z\left(q^2\right)=\int^1_0\int^{1-u}_0 \frac{dudv}{\Delta_Z}\Bigg( (3 u-1) \Delta_Z \log\left(\frac{\Delta_Z}{\mu^2}\right)+2 \left(m_q^2 (u-1)^3-2 m_Z^2 u\right)+2 q^2 u v
   (u+v-1)  \Bigg),
\end{equation}
and
\begin{equation}
\mathcal{V}_q^Z\left(q^2\right)=-\int^1_0\int^{1-u}_0 \frac{dudv}{\Delta_Z}   m_Z^2 (u-1) u,
\end{equation}
where $\Delta_Z=m_q^2 (u-1)^2+m_Z^2 u+q^2 v(u+v-1)$ and $\mu$ is the scale of dimensional regularization, which cancels out after integration.

For $q^2=0$, the last two loop functions give:
\begin{equation}
\mathcal{A}_q^Z(0)=\int_0^1du\frac{  u \left(2  u^2+r_{Z,q}^2
   (u-4) (u-1)\right)}{
    r_{Z.q}^2 (u-1)- u^2},
\end{equation}
and
\begin{equation}
\mathcal{V}_q^Z(0)=\int_0^1du\frac{ {g_V^q}^2 r_{Z,q}^2 (u-1) u^2}{
    r_{Z.q}^2 (u-1)- u^2}.
\end{equation}
As for the $W$ boson exchange contribution of diagram \ref{EWDiagrams}(b), it is given by [Eq. \eqref{Wexchangeat}]:
\begin{align}
\mathcal{F}_{qq'}^W\left(q^2\right)&=\int^1_0\int^{1-u}_0 \frac{dudv}{\Delta_W}\Bigg( -(1-3 u) \log \left( \frac{\Delta_W}{\mu^2}\right)\Delta_W-2m_{q^\prime}^2(u-1)^2+u\big(2m_q^2(u-1)^2\nonumber\\
&-m_W^2(u+3)+2 q^2 v(u+y-1) \big)\Bigg),
\end{align}
with the following result for $q^2=0$:
\begin{equation}
\mathcal{F}_{qq'}^W(0)=\int_0^1\frac{ u \left[u \left( (u-1)+r_{q^\prime , q}^2 (u+1)\right)+2
   r_{W , q}^2 (u-2) (u-1)\right]}{r_{W , q}^2 (u-1)-u
   \left( (u-1)+r_{q^\prime , q}^2\right)}du,
\end{equation}
where $\Delta_W= u \left(m_{q}^2 (u-1)+m_W^2\right)-m_{q^\prime}^2 (u-1)+q^2 v(u+v-1)$. Again this calculation was done via the unitary gauge.

Finally, for the Higgs contribution  [Eq. \eqref{Higgsexchangeat}] we obtain:
\begin{equation}
\mathcal{F}_q^h\left(q^2\right)=m_q^2\int_0^1\int_0^{1-u}\frac{ (u^2-1) }{m_q^2 (u-1)^2+u m_h^2
   + q^2 v(u+v-1)} dv du,
\end{equation}
which for $q^2=0$ reduces to
\begin{equation}
\mathcal{F}_q^h(0)=-\int^1_0 du \frac{(1+u)(1-u)^2}{(1-u)^2+ur_{h,q}^2}.
\end{equation}

\subsection{Passarino-Veltman results}
\label{PV}
We now present the above results in terms of Passarino-Veltman scalar integrals, where we use the standard notation for the two- and three-point scalar functions. Our calculation was done via a renormalizable linear $R_\xi$ gauge and the BFM to verify that the dependence on the $\xi$ gauge parameter drops out.   The loop functions are thus gauge independent and read

\begin{equation}
{\cal F}_q^{{\rm QCD}_1}\left(q^2\right)=\frac{m_q^2
   }{\eta^2(\|q\|,m_q)}\Big\{\mathbf{B}_0(0;m_q,m_q)-\mathbf{B}_0\left(q^2;m_q,m_q\right)+2\Big\},
\end{equation}
where we define $\eta(x,y)=\sqrt{x^2-4y^2}$. Also

\begin{align}
{\cal F}_q^{{\rm QCD}_2}\left(q^2\right)&=\frac{m_q^2}{\eta^4(\|q\|,m_q)} \Big\{\left(8 m_q^2+q^2\right)\left( \mathbf{B}_0(0;m_q,m_q)-\mathbf{B}_0\left(q^2;0,0\right)\right)\nonumber\\
   &-6 m_q^2 \left(q^2
   \mathbf{C}_0\left(m_q^2,m_q^2,q^2;0,m_q,0\right)-4\right)\Big\},
   \end{align}

\begin{equation}
\label{aQCD20}
\mathcal{F}_q^{{\rm QCD}_2}(0)=\frac{1}{2}\Big\{\mathbf{B}_0(0;m_q,m_q)+3\Big\},
\end{equation}

\begin{equation}
\mathcal{F}_q^{A}\left(q^2\right)=\frac{m_q^2}{\eta^2(\|q\|,m_q)}\Big\{\mathbf{B}_0\left(0;m_q,m_q\right)-\mathbf{B}_0\left(q^2;m_q,m_q\right)+2\Big\},
\end{equation}

\begin{align}
\mathcal{A}_q^{Z}\left(q^2\right)&=\frac{m_Z^2}{m_q^2\eta^4(\|q\|,m_q)} \Bigg\{\Big(q^2 \left(m_Z^2-8
   m_q^2\right)+10 m_q^2
   \left(2 m_q^2-m_Z^2\right)\Big) \mathbf{B}_0\left(m_q^2;m_q,m_Z\right)\nonumber \\
   &+\frac{m_q^2}{m_Z^2} \left(q^2 \left(9
   m_Z^2-2 m_q^2\right)+
   \left(8 m_q^4-24 m_q^2 m_Z^2+6 m_Z^4\right)\right)\mathbf{B}_0\left(q^2;m_q,m_q\right)\nonumber\\
   &+
   \frac{\left(2 m_q^2+m_Z^2\right) \eta^2(\|q\|,m_q)}{m_Z^2}\left(m_q^2\mathbf{B}_0(0;m_q,m_q)-m_Z^2
    \mathbf{B}_0(0;m_Z,m_Z)\right)\nonumber\\
   &+2\Big(2  q^2 \left(3 m_Z^2-7
   m_q^2\right) +2  q^4
   +3\left(8 m_q^4-6 m_q^2
   m_Z^2+m_Z^4\right)\Big)m_q^2
   \mathbf{C}_0\left(m_q^2,m_q^2,q^2;m_q,m_Z,m_q\right)\nonumber\\
   &+\frac{(4m_q^4-m_Z^4) \eta^2(\|q\|,m_q)}{m_Z^2}\Bigg\},
\end{align}

\begin{align}
\mathcal{V}_q^{Z}\left(q^2\right)&=\frac{m_Z^2}{m_q^2\eta^4(\|q\|,m_q)}
\Bigg\{\Big(q^2 \left(m_Z^2-2 m_q^2\right)
   +2 m_q^2 \left(4 m_q^2-5 m_Z^2\right)\Big)
   \mathbf{B}_0\left(m_q^2;m_q,m_Z\right)\nonumber\\
   &+
m_q^2\Big(2  \left(3 m_Z^2-2 m_q^2\right)
   + q^2 \Big)\mathbf{B}_0\left(q^2;m_q,m_q\right)+ \eta^2(\|q\|,m_q)\left(m_q^2\mathbf{B}_0(0;m_q,m_q)-m_Z^2
   \mathbf{B}_0(0;m_Z,m_Z)\right)\nonumber\\
   &+\Big(4  q^2
   +2
   \left(3 m_Z^2-8 m_q^2\right)\Big)
   m_q^2
   m_Z^2\mathbf{C}_0\left(m_q^2,m_q^2,q^2;m_q,m_Z,m_q\right)+(2m_q^2-m_Z^2)
   \eta^2(\|q\|,m_q)\Bigg\},
   \end{align}

\begin{align}
\mathcal{A}_q^{Z}(0)&=\frac{m_Z^2}{8m_q^4}\Bigg\{\Big(10 m_q^2-5 m_Z^2\Big)
   \mathbf{B}_0\left(m_q^2;m_q,m_Z\right)+\Big(3 m_Z^2 -14 m_q^2  \Big)\mathbf{B}_0(0;m_q,m_q)\nonumber\\
   &+2\Big(2 m_q^2 +
   m_Z^2 \Big)\mathbf{B}_0(0;m_Z,m_Z)+3\Big(m_Z^4
   -6 m_q^2 m_Z^2
   +8 m_q^4 \Big)
   \mathbf{C}_0\left(m_q^2,m_q^2,0;m_q,m_Z,m_q\right)\nonumber\\
   &+\frac{2}{m_Z^2} \left(m_Z^4-4m_q^4\right)\Bigg\},
\end{align}
\begin{align}
\mathcal{V}_q^{Z}(0)&=\frac{m_Z^2}{8m_q^4}
\Bigg\{\Big(4 m_q^2
   -5 m_Z^2\Big) \mathbf{B}_0\left(m_q^2;m_q,m_Z\right)+\Big(3 m_Z^2-4 m_q^2\Big) \mathbf{B}_0(0;m_q,m_q)\nonumber\\
  &+2 m_Z^2 \mathbf{B}_0(0;m_Z,m_Z)
   +\Big(3 m_Z^2
  -8 m_q^2\Big)m_Z^2
   \mathbf{C}_0\left(m_q^2,m_q^2,0;m_q,m_Z,m_q\right)+2 \left(m_Z^2-2
   m_q^2\right)\Bigg\},
\end{align}

\begin{align}
\mathcal{F}_{qq'}^{W}\left(q^2\right)&=\frac{1}{m_q^2\eta^4(\|q\|,m_q)}  \Big\{ \eta^2(\|q\|,m_q)
   \left(m_q^2+m_{q^\prime}^2+2 m_W^2\right)\left(m_{q^\prime}^2\mathbf{B}_0(0;m_{q^\prime},m_{q^\prime})
-m_W^2\mathbf{B}_0(0;m_W,m_W)\right)
   \nonumber\\
   &-\Big[q^2
   \left(m_q^4+m_q^2 \left(9 m_W^2-2
   m_{q^\prime}^2\right)+m_{q^\prime}^4+m_{q^\prime}^2 m_W^2-2 m_W^4\right)+2
   m_q^2 \Big(m_q^4+m_q^2 \left(4 m_{q^\prime}^2-9 m_W^2\right)\nonumber\\
   &-5
   \left(m_{q^\prime}^4+m_{q^\prime}^2 m_W^2-2 m_W^4\right)\Big)\Big]
   \mathbf{B}_0\left(m_q^2;m_{q^\prime},m_W\right)
+m_q^2 \Big[q^2
   \left(m_q^2-3 m_{q^\prime}^2+10 m_W^2\right)\nonumber\\
&+2 \left(m_q^4+m_q^2
   \left(6 m_{q^\prime}^2-11 m_W^2\right)-3 \left(m_{q^\prime}^4+m_{q^\prime}^2
   m_W^2-2 m_W^4\right)\right)\Big]
   \mathbf{B}_0\left(q^2;m_{q^\prime},m_{q^\prime}\right)\nonumber\\
   &-2 m_q^2
   \Big[m_q^2\left(m_q^4-m_q^2 \left(5 m_{q^\prime}^2+12 m_W^2\right)+\left(7 m_{q^\prime}^4-12 m_{q^\prime}^2 m_W^2+17 m_W^4\right)\right)
-3 \left(m_{q^\prime}^6-3 m_{q^\prime}^2
   m_W^4+2 m_W^6\right)
   \nonumber\\&-q^2
   \left(m_q^4-2 m_q^2 \left(m_{q^\prime}^2+4 m_W^2\right)+m_{q^\prime}^4-6
   m_{q^\prime}^2 m_W^2+8 m_W^4\right)-2 m_W^2 \left(q^2\right)^2\Big]
   \mathbf{C}_0\left(m_q^2,m_q^2,q^2;m_{q^\prime},m_W,m_{q^\prime}\right)\nonumber\\
   &+\eta^2(\|q\|,m_q)
   \left(m_q^2+m_{q^\prime}^2-m_W^2\right) \left(m_q^2+m_{q^\prime}^2+2
   m_W^2\right)\Big\},
   \end{align}

\begin{align}
\mathcal{F}_{qq'}^{W}(0)&=\frac{1}{ 8m_{q}^4}\Big\{2 m_W^2 \left(m_{q}^2+m_{q^\prime}^2+2 m_W^2\right)
   \mathbf{B}_0(0;m_W,m_W)+\Big[m_{q}^4+m_{q}^2 \left(4
   m_{q^\prime}^2-11 m_W^2\right)\nonumber\\
   &-5 m_{q^\prime}^4-7 m_{q^\prime}^2 m_W^2+6
   m_W^4\Big] \mathbf{B}_0(0;m_{q^\prime},m_{q^\prime})-\Big[m_{q}^4+m_{q}^2
   \left(4 m_{q^\prime}^2-9 m_W^2\right)\nonumber\\
   &-5 \left(m_{q^\prime}^4+m_{q^\prime}^2
   m_W^2-2 m_W^4\right)\Big]
   \mathbf{B}_0\left(m_{q}^2;m_{q^\prime},m_W\right)+\Big[12 m_{q}^2 m_W^2
   \left(m_{q}^2+m_{q^\prime}^2\right)\nonumber\\
   &-m_W^4 \left(17
   m_{q}^2+9 m_{q^\prime}^2\right)-\left(m_{q}^2-3 m_{q^\prime}^2\right)
   \left(m_{q}^2-m_{q^\prime}^2\right)^2+6 m_W^6\Big]
   \mathbf{C}_0\left(m_{q}^2,m_{q}^2,0;m_{q^\prime},m_W,m_{q^\prime}\right)\nonumber\\
   &-2
   \left(m_{q}^2+m_{q^\prime}^2-m_W^2\right) \left(m_{q}^2+m_{q^\prime}^2+2
   m_W^2\right)\Big\},
\end{align}

\begin{align}
\mathcal{F}_q^{h}\left(q^2\right)&=\frac{1}{\eta^4(\|q\|,m_q)} \Big\{\eta^2(\|q\|,m_q)\left(
   m_h^2\mathbf{B}_0(0;m_h,m_h)-m_q^2
   \mathbf{B}_0(0;m_q,m_q)\right)\nonumber\\
  & +\left(4m_q^2\eta^2(\|q\|,m_q)+m_h^2(10 m_q^2- q^2)\right) \mathbf{B}_0\left(m_q^2;m_q,m_h\right)-3m_q^2\left(\eta^2(\|q\|,m_q)-2
   m_h^2\right) \mathbf{B}_0\left(q^2;m_q,m_q\right)\nonumber\\
   &-6m_h^2 m_q^2\left(\eta^2(\|q\|,m_q)+m_h^2\right) \mathbf{C}_0\left(m_q^2,m_q^2,q^2;m_q,m_h,m_q\right)+(m_h^2-2m_q^2)\eta^2(\|q\|,m_q)\Big\},
\end{align}
and
\begin{align}
\mathcal{F}_q^{h}(0)&=-\frac{1}{8m_q^2}\Big\{\left(3 m_h^2-8 m_q^2\right) \mathbf{B}_0(0;m_q,m_q)+\left(8
   m_q^2-5 m_h^2\right) \mathbf{B}_0\left(m_q^2;m_q,m_h\right)\nonumber\\
   &+2
   m_h^2 \mathbf{B}_0(0;m_h,m_h)+3 m_h^2
   \left(m_h^2-4 m_q^2\right) \mathbf{C}_0\left(m_q^2,m_q^2,0;m_q,m_h,m_q\right)\nonumber\\
   &+2 (m_h^2-2m_q^2)\Big\}.
\end{align}

\subsection{Closed form results}\label{analy}
We also present the explicit solutions for the two-point scalar functions in terms of closed form functions. Below $C_0\left(m_i^2,m_j^2,q^2;m_k,m_l,m_n\right)$ stands for a three-point Passarino-Veltman scalar function.

\begin{equation}
{\mathcal F}_q^{{\rm QCD}_1}\left(q^2\right)=-\frac{m_q^2}{\|q\|\eta(\|q\|,m_q)}\log \left(\frac{\|q\|\eta(\|q\|,m_q)+2
   m_q^2-q^2}{2 m_q^2}\right),
\end{equation}

\begin{align}
{\mathcal F}_q^{{\rm QCD}_2}\left(q^2\right)&=-\frac{m_q^2}{\eta^4(\|q\|,m_q)} \Bigg\{q^2 \Bigg[6 m_q^2
   C_0\left(m_q^2,m_q^2,q^2;0,m_q,0\right)+\log \left(\frac{m_q^2}{q^2}\right)+2+i\pi\Bigg]\nonumber \\
   &+8 m_q^2
   \Bigg[\log \left(\frac{m_q^2}{q^2}\right)-1+i\pi\Bigg]\Bigg\},
\end{align}

\begin{equation}
\label{QCD_2pole}
{\mathcal F}_q^{{\rm QCD}_2}(0)=  \frac{1}{2} \Bigg\{\frac{1}{\epsilon }+\log \left(\frac{\mu
   ^2}{m_q^2}\right)+3\Bigg\},
\end{equation}

\begin{align}
{\mathcal F}_q^{A}\left(q^2\right)&=-\frac{m_q^2}{\|q\|\eta(\|q\|,m_q)}\log \left(\frac{\|q\|\eta(\|q\|,m_q)+2 m_q^2-q^2}{2
   m_q^2}\right),
\end{align}

\begin{align}
{\mathcal A}_q^{Z}\left(q^2\right)&=\frac{m_Z^2}{2 m_q^2\eta^4(\|q\|,m_q)}
  \Bigg\{
\Bigg(20 m_Z \left(2 m_q^2-m_Z^2\right)+\frac{2 m_Z \left(m_Z^2-8
   m_q^2\right) q^2 }{m_q^2}\Bigg)  \eta(m_Z,m_q) \log \left(\frac{m_Z+\eta(m_Z,m_q)}{2
   m_q}\right)\nonumber\\&+
  2\Bigg(\frac{ \left(9 m_Z^2-2
   m_q^2\right) }{m_Z^2}+\frac{ \left(8 m_q^4-24
   m_Z^2 m_q^2+6 m_Z^4\right) }{m_Z^2 q^2} \Bigg)m_q^2\|q\|\eta(\|q\|,m_q)\log \left(\frac{2 m_q^2-q^2+\|q\|\eta(\|q\|,m_q)}{2 m_q^2}\right)\nonumber\\&+\Bigg(2 \left(8 m_q^4+14 m_Z^2 m_q^2-5 m_Z^4\right) +\frac{\left(-4 m_q^4-10 m_Z^2 m_q^2+m_Z^4\right) q^2}{m_q^2}\Bigg)\log
   \left(\frac{m_q^2}{m_Z^2}\right)\nonumber\\
   &+
   \Big(8 \left(q^2\right)^2 +12 \left(8 m_q^4-6 m_Z^2 m_q^2+m_Z^4\right)
   +8 \left(3 m_Z^2-7 m_q^2\right) q^2\Big)
   m_q^2C_0\left(m_q^2,m_q^2,q^2;m_q,m_Z,m_q\right) \nonumber
   \\&+2 \left(2 m_q^2+m_Z^2\right)\eta^2(\|q\|,m_q)
   \Bigg\},
   \end{align}

\begin{align}
{\mathcal V}_q^{Z}\left(q^2\right)&=\frac{m_Z^2}{2 m_q^2\eta^4(\|q\|,m_q)}\Bigg\{
   2m_Z^2\Big( \left(3 m_Z^2-8 m_q^2\right)
   +2 q^2\Big)
   m_q^2C_0\left(m_q^2,m_q^2,q^2;m_q,m_Z,m_q\right) \nonumber\\
   &+\Bigg(\frac{4 \left(3 m_Z^2-2 m_q^2\right) }{q^2}+2\Bigg) m_q^2\|q\|\eta^2(\|q\|,m_q) \log
   \left(\frac{2 m_q^2-q^2+\|q\|\eta(\|q\|,m_q)}{2 m_q^2}\right)\nonumber \\
   &+2 m_Z^2 \eta(\|q\|,m_q)+\Bigg(2
   m_Z^2 \left(8 m_q^2-5 m_Z^2\right)+\frac{m_Z^2 \eta^2(m_Z,m_q)q^2 }{m_q^2}\Bigg) \log
   \left(\frac{m_q^2}{m_Z^2}\right)\nonumber\\
   &+\Bigg(4 m_Z \left(4 m_q^2-5 m_Z^2\right)+\frac{2 m_Z
   \left(m_Z^2-2 m_q^2\right) q^2 }{m_q^2}\Bigg)
   \eta(m_Z,m_q) \log \left(\frac{m_Z+\eta(m_Z,m_q)}{2
   m_q}\right)\Bigg\},
\end{align}

\begin{align}
{\mathcal A}_q^{Z}(0)&=\frac{m_Z^2}{2
   m_q^6}\Bigg\{-\frac{m_q^2 \left(m_q^2-2 m_Z^2\right)
   \left(2 m_q^2-m_Z^2\right)}{m_Z^2}+\left(-2 m_q^4+4 m_q^2
   m_Z^2-m_Z^4\right) \log \left(\frac{m_q^2}{m_Z^2}\right)\nonumber\\
   &-\frac{2\left(8 m_q^4-6 m_q^2 m_Z^2+m_Z^4\right) m_Z}{\eta(m_Z,m_q)}
   \log \left(\frac{m_Z+\eta(m_Z,m_q)}{2 m_q
   }\right)\Bigg\},
   \end{align}

\begin{align}
{\mathcal V}_q^{Z}(0)&=\frac{m_Z^2}{2
   m_q^6}\Bigg\{m_q^2 \left(m_q^2-2
   m_Z^2\right)-m_Z^2 \left(m_Z^2-2 m_q^2\right) \log
   \left(\frac{m_q^2}{m_Z^2}\right)\nonumber\\
   &-\frac{2
   \left(2 m_q^4-4 m_q^2 m_Z^2+m_Z^4\right)}{\eta(m_Z,m_q)}
   \log \left(\frac{\eta(m_Z,m_q)+m_Z}{2 m_q}\right)\Bigg\},
   \end{align}

\begin{align}
{\mathcal F}_{qq'}^{W}\left(q^2\right)&= \frac{1}{2 m_q^2\eta^4(\|q\|,m_q)} \Bigg\{-4 m_q^2 \Big(m_q^6-m_q^4 \left(5
   m_{q^\prime}^2+12 m_W^2\right)+m_q^2 \left(7 m_{q^\prime}^4-12 m_{q^\prime}^2
   m_W^2+17 m_W^4\right)\nonumber\\
   &-q^2 \left(m_q^4-2 m_q^2
   \left(m_{q^\prime}^2+4 m_W^2\right)+m_{q^\prime}^4-6 m_{q^\prime}^2 m_W^2+8
   m_W^4\right)-3 \left(m_{q^\prime}^6-3 m_{q^\prime}^2 m_W^4+2
   m_W^6\right)\nonumber\\
   &-2 m_W^2 \left(q^2\right)^2\Big)
   C_0\left(m_q^2,m_q^2,q^2;m_{q^\prime},m_W,m_{q^\prime}\right)+2 \eta^2(\|q\|,m_q) \big(m_q^4+3 m_q^2
   m_W^2-m_{q^\prime}^4-m_{q^\prime}^2 m_W^2+2 m_W^4\big)\nonumber\\
   &+\frac{2
   m_q^2 }{q^2}\|q\|\eta(\|q\|,m_q')  \Big(2 \left(m_q^4+m_q^2\left(6 m_{q^\prime}^2-11 m_W^2\right)-3 \left(m_{q^\prime}^4+m_{q^\prime}^2
   m_W^2-2 m_W^4\right)\right)\nonumber\\
   &+q^2
   \big(-3 m_{q^\prime}^2+m_q^2+10 m_W^2\big)\Big)\log \left(\frac{\|q\|\eta(\|q\|,m_q')+2 m_{q^\prime}^2-q^2}{2 m_{q^\prime}^2}\right)\nonumber\\
   &-\frac{2 }{m_q^2}\sqrt{((m_q - m_q')^2 - m_W^2) ((m_q + m_q')^2 - m_W^2)}
   \Big(q^2 \big(m_q^4+m_q^2 \left(9 m_W^2-2m_{q^\prime}^2\right)+m_{q^\prime}^4+m_{q^\prime}^2 m_W^2\nonumber\\
   &-2 m_W^4\big)+2
   m_q^2 \left(m_q^4+m_q^2 \left(4 m_{q^\prime}^2-9 m_W^2\right)-5
   \left(m_{q^\prime}^4+m_{q^\prime}^2 m_W^2-2 m_W^4\right)\right)\Big)\nonumber\\
   & \times \log
   \left(\frac{\sqrt{((m_q - m_q')^2 - m_W^2) ((m_q + m_q')^2 - m_W^2)}+m_{q^\prime}^2-m_q^2+m_W^2}{2 m_{q^\prime} m_W}\right)\nonumber\\
   &-\frac{1}{m_q^2} \Big(q^2 \big(m_q^6-3
   m_q^4 \left(m_{q^\prime}^2-4 m_W^2\right)+m_q^2 \left(3 m_{q^\prime}^4-8
   m_{q^\prime}^2 m_W^2+11 m_W^4\right)\nonumber\\
   &-m_{q^\prime}^6+3 m_{q^\prime}^2 m_W^4-2
   m_W^6\big)+2 m_q^2 \big(m_q^6+3 m_q^4 \left(m_{q^\prime}^2-4
   m_W^2\right)+m_q^2 \left(-9 m_{q^\prime}^4+4 m_{q^\prime}^2 m_W^2-7
   m_W^4\right)\nonumber\\
   &+5 \left(m_{q^\prime}^6-3 m_{q^\prime}^2 m_W^4+2
   m_W^6\right)\big)\Big)\log
   \left(\frac{m_{q^\prime}^2}{m_W^2}\right)\Bigg\},
\end{align}

\begin{align}
{\mathcal F}_{qq'}^{W}(0)&=\frac{1}
   {2m_q^6} \Bigg\{m_q^2 \Big(-m_q^4-m_q^2 \left(3
   m_{q^\prime}^2-4 m_W^2\right)+2 \left(m_{q^\prime}^4+m_{q^\prime}^2 m_W^2-2
   m_W^4\right)\Big)\nonumber\\
   &-\left(m_q^4 m_{q^\prime}^2+m_q^2 \left(-2
   m_{q^\prime}^4+2 m_{q^\prime}^2 m_W^2-3 m_W^4\right)+m_{q^\prime}^6-3 m_{q^\prime}^2
   m_W^4+2 m_W^6\right) \log
   \left(\frac{m_{q^\prime}^2}{m_W^2}\right)\nonumber\\
   &-\frac{2}{\sqrt{((m_q - m_q')^2 - m_W^2) ((m_q + m_q')^2 - m_W^2)}} \Big(m_q^6
   m_{q^\prime}^2+m_q^4 \left(-3 m_{q^\prime}^4+m_{q^\prime}^2 m_W^2+3
   m_W^4\right)\nonumber\\
   &+m_q^2 \left(3 m_{q^\prime}^6-2 m_{q^\prime}^4 m_W^2+4
   m_{q^\prime}^2 m_W^4-5 m_W^6\right)-\left(m_{q^\prime}^2-m_W^2\right)^3
   \left(m_{q^\prime}^2+2 m_W^2\right)\Big) \nonumber\\
   &\times\log\left(\frac{\sqrt{((m_q - m_q')^2 - m_W^2) ((m_q + m_q')^2 - m_W^2)}+m_{q^\prime}^2-m_q^2+m_W^2}{2 m_{q^\prime} m_W}\right)\Bigg\},
\end{align}

\begin{align}
{\mathcal F}_q^{h}\left(q^2\right)&=-\frac{1}{2m_q^2 \eta^4(\|q\|,m_q)} \Bigg\{12 m_q^4 m_h^2 \left(\eta(\|q\|,m_q)+m_h^2\right)
   C_0\left(m_q^2,m_q^2,q^2;m_q,m_h,m_q\right)\nonumber\\
   &+2 m_q^2 m_h^2 \eta(\|q\|,m_q)+\frac{6 m_q^4}{q^2} \|q\|\eta(\|q\|,m_q) \left(\eta^2(\|q\|,m_q)+2m_h^2\right) \log \left(\frac{\|q\|\eta(\|q\|,m_q)+2
   m_q^2-q^2}{2 m_q^2}\right)\nonumber\\
   &+m_h^2 \left(24 m_q^4+q^2 \left(m_h^2-6 m_q^2\right)-10 m_q^2 m_h^2\right) \log \left(\frac{m_q^2}{m_h^2}\right)\nonumber\\
   &-2 m_h\eta(m_h,m_q) \left(-16 m_q^4-q^2
   \eta(m_h,m_q)+10 m_q^2 m_h^2\right) \log
   \left(\frac{\eta(m_h,m_q)+m_h}{2 m_q
   }\right)\Bigg\},
\end{align}
and
\begin{align}
{\cal F}_q^{h}(0)&=-\frac{1}{2m_q^4} \Bigg\{m_q^2(3 m_q^2-2  m_h^2)+2 m_h
   \left(m_q^2-m_h^2\right) \eta(m_h,m_q) \log
   \left(\frac{\eta(m_h,m_q)+m_h}{2 m_q}\right)\nonumber\\
   &+m_h^2\left(3 m_q^2-m_h^2\right) \log \left(\frac{m_q^2}{m_h^2}\right)\Bigg\}.
   \end{align}

\twocolumn

\end{document}